\documentclass[useAMS,usenatbib]{mn2e}

\usepackage{graphicx,amssymb}

\title[Accreting magnetars]{Accreting magnetars: linking ultraluminous X-ray pulsars and the slow pulsation X-ray pulsars}
\author[Tong \& Wang]{H. Tong$^{1}$\thanks{E-mail:htong\_2005@163.com} and W. Wang$^{2,3}$\thanks{E-mail:wangwei2017@whu.edu.cn}\\
$^{1}$School of Physics and Electronic Engineering, Guangzhou University, Guangzhou 510006, China\\
$^{2}$School of Physics and Technology, Wuhan University, Wuhan 430072, China \\
$^{3}$National Astronomical Observatories, Chinese Academy of Sciences, Beijing 100012, China}
\begin{document}

\date{Accepted XXXX. Received XXXX; in original form XXXX}

\pagerange{\pageref{firstpage}--\pageref{lastpage}} \pubyear{2018}

\maketitle

\label{firstpage}

\begin{abstract}

Possible manifestations of accreting magnetars are discussed. It is shown that the four ultra-luminous X-ray
pulsars can be understood in the accreting low magnetic field magnetar scenario. The NGC300 ULX1 pulsar may have a higher
dipole magnetic field than other sources.
General constraint on their mass accretion rate confirmed their super-Eddington nature.
Lower limits on their beaming factor are obtained. They do not seem to have strong beaming.
The duty cycle of the ULX burst state can also be constrained by their timing observations.
ULX pulsars may be in accretion equilibrium in the long run. During the outburst, they
will spin up, and run from the previous equilibrium state to the new equilibrium state.
It is proposed that the slowest puslation X-ray pulsar AX J1910.7+0917
may be an accreting magnetar with a low mass accretion rate. ULX pulsars, slow pulsation X-ray pulsars may all be accreting magnetars with different accretion rates.
Seven possible signatures of an accreting magnetar are summarized.

\end{abstract}


\section{Introduction}

Magnetars form a special kind of pulsars. They may have magnetic field as high as $10^{15} \,\rm G$
(Duncan \& Thompson 1992; Kaspi \& Beloborodov 2017).  Strong magnetic field powers magnetar's
 persistent emissions and bursts. Observationally, some magnetars with low dipole magnetic field were found
 (Rea et al. 2010; Zhou et al. 2014). Their dipole magnetic field can be several times $10^{12} \,\rm G$.
 These low magnetic field magnetars may be old magnetars (Turolla et al. 2011; Tong \& Xu 2012; Dall'Osso et al. 2012).
 From the early time of magnetar researches, one open question is: where are accreting magnetars (Woods \& Thompson 2006)?
 It is expected that there are some accreting magnetars in binary systems.
 The problem is: how to find the signature of an accreting magnetar from the zoo of accreting neutron stars?

 Previously, it is speculated that magnetar-like bursts and a hard X-ray tail above $100\,\rm keV$ are the two
 observational signatures of accreting magnetars (Tong \& Wang 2014). Ultraluminous X-ray (ULX) pulsars are accreting neutron star
 with X-ray luminosity as high as $10^{40} \,\rm erg\,s^{-1}$ or higher (Bachetti et al. 2014).
 It is pointed out that ULX pulsars may be another manifestation of accreting magnetars (Eksi et al. 2015; Tong 2015; Dall'Osso et al. 2015; Lyutikov 2014; Tsygankov et al. 2016). The ultrastrong magnetic field may be responsible for the super-Eddington luminosity\footnote{There are also models using normal neutron stars for ULX pulsars (e.g., King et al. 2017; Christodoulou et al. 2017, 2018).}. However, different authors obtain different estimates of the neutron star's dipole magnetic field, ranging from $\gtrsim 10^{12}$ G (Dall'Osso et al. 2015, Lyutikov 2014) up to $\sim 10^{15}$ G (Tsygankov et al. 2016).
  Since old magnetars are more likely to be low magnetic field magnetars, it is proposed that
 the ULX pulsars may be accreting low magnetic field magnetars (Tong 2015).
 Later discovery of more ULX pulsars are consistent with the accreting low magnetic field magnetar scenario
 (Israel et al. 2017a,b).

If ULX pulsars are accreting magnetars  with a very high mass accretion rate, then some accreting magnetars
with relative low mass accretion rates are also expected.  Accretion systems are often in the accretion equilibrium state (Bhattacharya \& van den Heuvel 1991; Ho et al. 2014). The equilibrium period of an accreting neutron star depends on the stellar magnetic field and the mass accretion rate: $P_{\rm eq} \propto B^{6/7} \dot{M}^{-3/7}$. The rotational periods of ULX pulsars are about several seconds. Then, accreting magnetars with low mass accretion rates may have long rotational periods. In this respect, the slowest pulsation X-ray pulsar AX J1910.7$+$0917 with a rotational period of $3.6\times 10^4 \,\rm s$ may also be an accreting magnetar candidate (Sidoli et al. 2017). Other slow pulsation X-ray pulsars with rotational periods about
$10^3 \,\rm s$ or longer are also proposed to be accreting magnetar candidates (Wang 2009, 2011; Reig et al. 2012; Yang et al. 2017). Therefore, a variety of accreting neutron stars may be linked together in the accreting magnetar scenario.

Detailed calculation for the four ULX pulsars in the accreting low magnetic field magnetar model is presented in
section 2. The merit of understanding the slowest X-ray pulsar in the magnetar scenario is discussed in section 3.
Discussions are given in section 4.

\section{ULX pulsars as accreting low magnetic field magnetars}

\subsection{Three aspects of ULX pulsar observations}

From the first discovery of Bachetti et al. (2014), four ULX pulsars are identified up to now (Bachetti et al. 2014; F$\rm \ddot{u}$rst  et al. 2016; Israel et al. 2017a,b; Carpano et al. 2018). From these four sources, more can be learned about ULX pulsars.

\begin{enumerate}
\item
Pulse profile.  All four sources exhibit near sinusoidal pulse profiles with broad pulse features. This means that ULX pulsars are not likely to have a strong beaming. At the same time, a near sinusoidal profile requires that ULX pulsars should have a moderate
degree of beaming. A typical beaming factor of $b=0.2$ may be chosen (Feng \& Soria 2011).
For a given source, its true X-ray luminosity is related to its isotropic X-ray luminosity as
\begin{equation}
L_{\rm x} = b\, L_{\rm x,iso} =b\, L_{\rm x,iso,40} \times 10^{40} \rm \ erg \ s^{-1},
\end{equation}
where $L_{\rm x,iso,40}$ is the isotropic X-ray luminosity in units of $10^{40} \rm \ erg \ s^{-1}$.
The mass accretion rate is related to the true X-ray luminosity as
\begin{equation}
L_{\rm x} = \eta \dot{M} c^2,
\end{equation}
where $\eta$ is the energy conversion efficiency for accretion onto the neutron star. Typical values of $\eta=0.1$ can be used (Frank et al. 2002).

\item Super-Eddington luminosity.
The isotropic X-ray luminosity of  ULX pulsars can be as high $10^{41} \,\rm erg \, s^{-1}$ (Israel et al. 2017b).
With a moderate beaming factor,
the neutron star's true X-ray luminosity can be 100 times the traditional Eddington luminosity. This confirms the super-Eddington
nature of ULX pulsars. In the accreting magnetar scenario, a magnetic field strength $\ge 10^{14} \,\rm G$
will result in a critical luminosity about $10^{41} \,\rm erg \, s^{-1}$ (Paczynski 1992; Mushtukov et al. 2015).
Therefore, the super-Eddington luminosity of ULX pulsars can be safely understood in the accreting
magnetar scenario.

\item Rotational behavior.
All the four ULX pulsars have a variable luminosity. The X-ray luminosity in the high state can be one hundred times higher than the X-ray luminosity in the low state. When assuming accretion equilibrium of the neutron star, there may be ambiguities in determining which luminosity corresponds to the equilibrium state. Similar ambiguities also exist
when assuming transition between the accretion phase and the propeller phase due to the sparse of observations (Tsygankov et al. 2016; Israel et al. 2017a,b). Both of the above two assumptions will result in a magnetospheric radius equal to the corotation radius, which can be used to deterimine the dipole magnetic field of the central neutron star.
Another way is to study the accretion torque and compare it with the the spin-up behavior (Eksi et al. 2015; Dall'Osso et al. 2015;  Lyutikov 2014). In this way, the dipole magnetic field can also be determined. Whether the neutron star is in accretion equilibrium can be determined thereafter.

\end{enumerate}

\subsection{Calculations of the dipole magnetic field}
\label{section_B_calculation}

For M82 X-2, its pulsation period is $P=1.37 \rm \ s$, typical period derivative $\dot{P} =-2 \times 10^{-10} \rm s\  s^{-1}$, typical isotropic X-ray luminosity when the pulsation is detected $L_{\rm x,iso} \approx 10^{40} \rm \ erg \ s^{-1}$ (Bachetti et al. 2014).
Following the above discussions, its true X-ray luminosity is
\begin{equation}
L_{\rm x} = 2\times 10^{39} \left(  \frac{b}{0.2} \right) L_{\rm x,iso,40} \rm \ erg \ s^{-1}.
\end{equation}
For M82 X-2, its mass accretion rate is
\begin{equation}
\dot{M} =2.2 \times 10^{19} \, \left(  \frac{b}{0.2} \right) \left(  \frac{\eta}{0.1} \right)^{-1} L_{\rm x,iso,40} \rm \ g \ s^{-1}.
\end{equation}
The spin-up of the neutron star is governed by
\begin{equation}\label{eqn_spinup}
I \dot{\Omega} =- 2 \pi I \frac{\dot{P}}{P^2} = \dot{M} \sqrt{G M R_{\rm A}},
\end{equation}
where $I$ and $M$ are the neutron star moment of inertia and mass, respectively, $R_{\rm A}$ is the Alfv$\rm \acute{e}$n radius.
The Alfv$\rm \acute{e}$n radius is determined by the dipole magnetic moment and mass accretion rate\footnote{The magnetosphere radius is assumed to be equal to the Alfv$\rm \acute{e}$n radius. A commonly employed factor of $0.5$ is not included, because the
accretion disk for ULX pulsars may become thick and the geometry may be similar to the spherical case (King et al. 2017; Walton et al. 2018a).} (Shapiro \& Teukolsky 1983; Lai 2014)
\begin{equation}\label{eqn_Alfven radius}
 R_{\rm A} = 3.2\times 10^8 \, \mu_{30}^{4/7} M_1^{-1/7} \dot{M}_{17}^{-2/7} \rm\ cm,
\end{equation}
where $\mu =1/2 B_{\rm p} R^3$ is the magnetic dipole moment of the neutron star, $\mu_{30}$ is the dipole moment in units of $10^{30} \,\rm G\, cm^3$, $M_1$ is the neutron star mass in units of solar masses, $\dot{M}_{17}$ is the mass accretion rate in units of $10^{17} \,\rm g\, s^{-1}$, $B_{\rm p}$ is the polar dipole magnetic field\footnote{The dipole magnetic field at the equator is two times smaller than the polar magnetic field (Shapiro \& Teukolsky 1983).}, and $R$ is the neutron star radius. For ULX pulsars, radiation pressure may be significant at the magnetospheric radius. This will change the corresponding expression of the Alfv$\rm \acute{e}$n radius (Dall'Osso et al. 2016; Xu \& Li 2017). At the same time, the conclusions are only slightly modified when the radiation pressure are included (Dall'Osso et al. 2016; Xu \& Li 2017).

A general accretion torque are employed in some studies (Eksi et al. 2015; Dall'Osso et al. 2015; Lyutikov 2014).
It consists of both the matter torque (equation(\ref{eqn_spinup})) and the disk magnetosphere coupling. However, as discussed in the Ghosh \& Lamb model (Ghosh \& Lamb 1979; Ghosh 1995; Lai 2014), the accretion torque will be dominated
by the matter torque for slow rotators (i.e. the magnetospheric radius is much smaller than the corotation radius, or a fastness parameter much smaller than one). ULX pulsars tends to be in the high luminosity state when pulsations are detected.
During the outburst, the luminosity of ULX pulsars can vary by a factor 10 (Israel et al. 2017a,b). The quiescent luminosity is more than 100 times smaller than the peak luminosity (Tsygankov et al. 2016; Dall'Osso et al. 2016).
Denoting the peak luminosity as $L_{\rm peak}$ when pulsations are detected, the minimum luminosity in the outburst is about
$0.1 L_{\rm peak}$. The long term averaged luminosity may be smaller than $0.01 L_{\rm peak}$. If ULX pulsars are in accretion equilibrium in the long run, then it is possible that they are in the slower rotator regime in the high luminosity state (especially in the peak luminosity state).

From equation (\ref{eqn_spinup}), the dipole magnetic field of M82 X-2 can be determined
\begin{equation}
B_{\rm p} =2.1 \times 10^{10} \left( \frac{b}{0.2}  \right)^{-3}  \left(  \frac{\eta}{0.1}  \right)^{3} L_{\rm x,iso,40}^{-3}  \rm \ G.
\end{equation}
During the calculations, typical values of neutron star mass $M=1.4 \, M_{\odot}$, and moment of inertia $I =10^{45} \,\rm g\, cm^2$ are adopted\footnote{The dependence on neutron star mass and moment of inertia is: $B_{\rm p} \propto (M/1.4M_{\odot})^{-3/2} (I/10^{45} \ \rm g \ cm^2)^{7/2}.$}. There are still parameter space that the dipole magnetic field of M82 X-2 can be as high as $10^{12} \ \rm G$. For example, for an energy conversion efficiency of $\eta =0.3$, the dipole magnetic field of M82 X-2 is about $0.6\times 10^{12} \,\rm G$. And a beaming factor two times smaller $b=0.1$ will enhance the result by a factor of eight.

The equilibrium period is reached when the magnetospheric radius is equal to the corotation radius. It is determined by the long term averaged mass accretion rate (Tong 2015)
\begin{equation}\label{eqn_Peq}
P_{\rm eq} = 3.1 \mu_{30}^{6/7} M_{1}^{-5/7} \dot{M}_{\rm ave, 17}^{-3/7} \rm \ s,
\end{equation}
where $\dot{M}_{\rm ave,17}$ is the long term averaged mass accretion rate in units of $10^{17} \,\rm g \,s^{-1}$. 
Therefore, M82 X-2 (as one example of the ULX pulsars) may be in accretion equilibrium provided that its dipole magnetic field is about $10^{12} \ \rm G$ and its long term averaged mass accretion rate is about $10^{17} \ \rm g \ s^{-1}$. However, the term ``average mass accretion rate'' is ill-defined and lacks physical meaning. A physical treatment is presented in section 2.5.

A surface magnetic field strength about or higher than  $10^{14} \,\rm G$ is required for the super-Eddington luminosity (e.g., as high as $10^{41} \ \rm erg \ s^{-1}$) of ULX pulsars (Paczynski 1992; Mushtukov et al. 2015).  While the spin-up behavior implies only a relatively low surface dipole field. This corresponds to the magnetic field configuration of low magnetic field magnetars (Rea et al. 2010; Tiengo et al. 2013). Therefore, an accreting low magnetic field magnetar may explain both the luminosity and rotational behavior of ULX pulsars (Tong 2015; Israel et al. 2017a,b):
the magnetar strength total magnetic field ($>10^{14} \,\rm G$) is responsible for the super-Eddington luminosity, and the lower dipole magnetic field ($\sim 10^{12} \,\rm G$) is responsible for the rotational behaviors.

Later observations of three more ULX pulsars are consistent with the accreting low magnetic field magnetar scenario\footnote{For Swift J0243.6+6124, which is a candidate ULX pulsar in the Galaxy, similar calculations showed that it has dipole magnetic field of $2.7\times 10^{13} \rm \ G$. The commonly reported magnetic field at the equator is two time smaller, about $1.3\times 10^{13} \rm \ G$. It is consistent with the result of Doroshenko et al. (2018) and justifies our slow rotator assumption.}.
\begin{itemize}
\item
For the ULX pulsar NGC7793 P13, its pulsation period is $P=0.42 \rm\ s$, long term spin-up rate $\dot{P} =-3.5\times 10^{-11} \rm\  s\ s^{-1}$, typical isotropic X-ray luminosity when pulsation is detected $L_{\rm x,iso} =10^{40} \rm\ erg \ s^{-1}$ (F$\rm \ddot{u}$rst et al. 2016; Israel et al. 2017a). Similar to the above calculations for M82 X-2, the dipole magnetic field for NGC7793 P13 is about:
$B_{\rm p} =2\times 10^{11} (b/0.2)^{-3} (\eta/0.1)^3 L_{\rm x,iso,40}^{-3} \rm\ G$.

\item
For NGC5907 ULX pulsar, its pulsation period is $P =1.137 \rm\ s$, period derivative in the high state $\dot{P} =-5\times 10^{-9} \rm \ s \ s^{-1}$, its isotropic X-ray luminosity can reach as high as $2\times 10^{41} \rm\ erg \ s^{-1}$ (Israel et a. 2017b). Similar to the above calculations, the dipole magnetic field of NGC5907 ULX pulsar is about:
$B_{\rm p} =6\times 10^{12} (b/0.2)^{-3} (\eta/0.1)^3 L_{\rm x,iso,41}^{-3} \rm\ G$, where $L_{\rm x,iso,41}$
is the star's isotropic X-ray luminosity in units of $10^{41} \,\rm erg \, s^{-1}$.

\item
For NGC300 ULX1 pulsar, it pulsation period is $P= 31.6 \rm \ s$, period derivative $\dot{P} =-5.56 \times 10^{-7} \rm s \ s^{-1}$, and typical X-ray luminosity $L_{\rm x, iso} = 4.7\times 10^{39} \rm \ erg \ s^{-1}$ (Carpano et al. 2018). Similar calculations showed that the dipole magnetic field of NGC300 ULX1 pulsar is about:
$B_{\rm p} =6.7\times 10^{13} (b/0.2)^{-3} (\eta/0.1)^3 (L_{\rm x,iso}/4.7\times 10^{39} {\rm erg \ s^{-1}})^{-3} \rm\ G$.
It has the highest dipole magnetic field among the four ULX pulsars.
\end{itemize}
The main parameters of the four ULX pulsars are summarized in table \ref{tab_four_ULX_pulsars}.

\begin{table*}
\caption{The main parameters of four ULX pulsars: the observed luminosity $L_{\rm x,iso}$, period $P$,
period derivative $\dot{P}$,
and the derived dipole magnetic field $B_{\rm p}$ assuming $b=0.2$ and $\eta=0.1$.}
\begin{center}\label{tab_four_ULX_pulsars}
\begin{tabular}{l c c c c l}
\hline \hline
ULX name & $L_{\rm x,iso}$ (erg s$^{-1}$) & $P$ (s) & $\dot {P}$ (s/s)  & $B_{\rm p}$ (G) & References \\
\hline
M82 X-2 & $10^{40}$ & 1.37 & -2$\times 10^{-10}$ & 2.1$\times 10^{10}$ & Bachetti et al. 2014 \\
NGC7793 P13 & $10^{40}$ & 0.42 & -3.5$\times 10^{-11}$ & 2$\times 10^{11}$
& F$\rm \ddot{u}$rst et al. 2016; Israel et al. 2017a  \\
NGC5907 ULX &  $2\times 10^{41}$ & 1.137 & -5$\times 10^{-9}$  & 6$\times 10^{12}$ & Israel et al. 2017b  \\
NGC300 ULX1 & $4.7\times 10^{39}$ &  31.6 & -5.56$\times 10^{-7}$ &  6.7$\times 10^{13}$ &  Carpano et al. 2018   \\
\hline
\end{tabular}
\end{center}
\end{table*}

\subsection{Correlations between $P$, $\dot{P}$ and $L_{\rm x}$}

\begin{figure}
\begin{minipage}{0.45\textwidth}
 \includegraphics[width=0.95\textwidth]{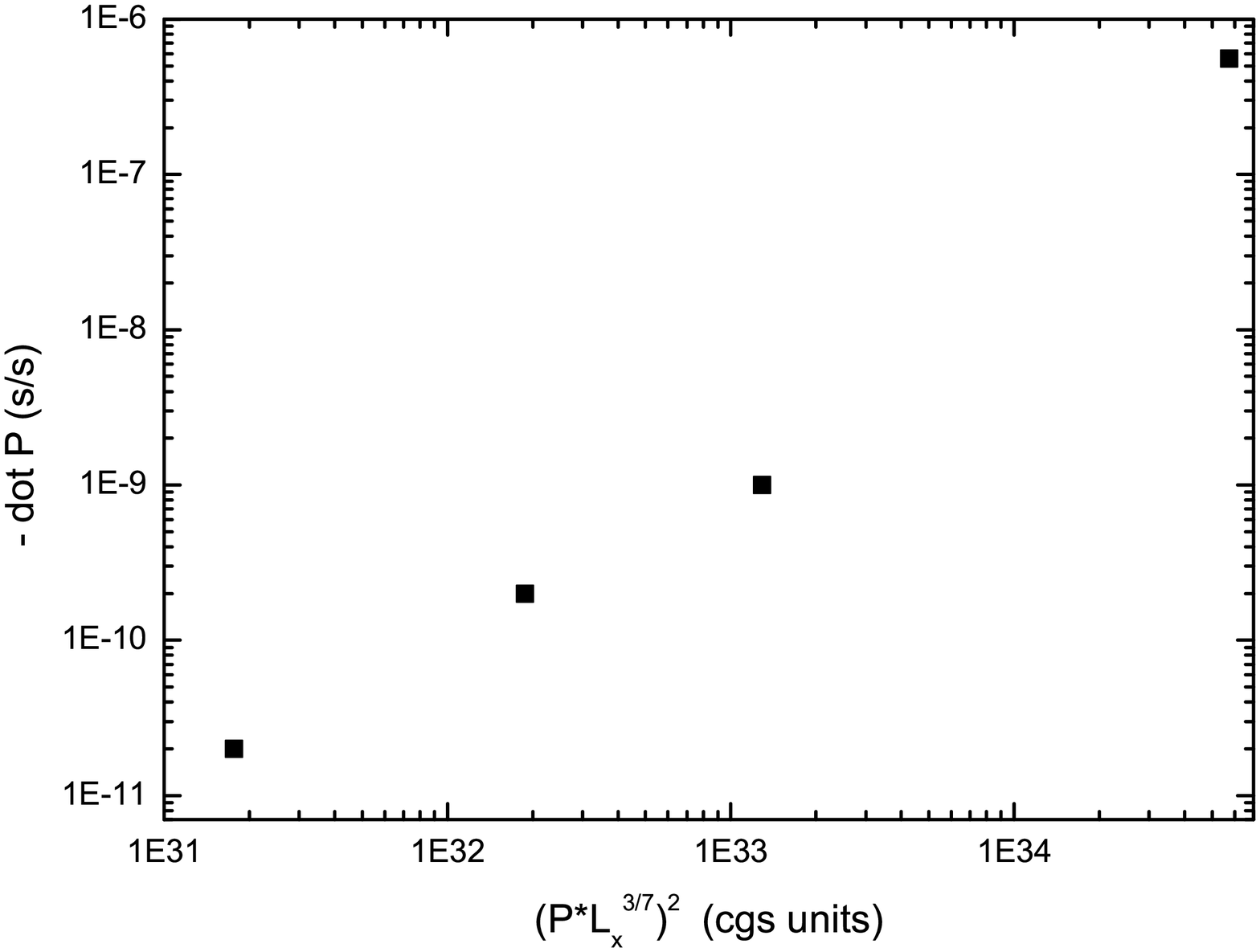}
\end{minipage}
\begin{minipage}{0.45\textwidth}
 \includegraphics[width=0.95\textwidth]{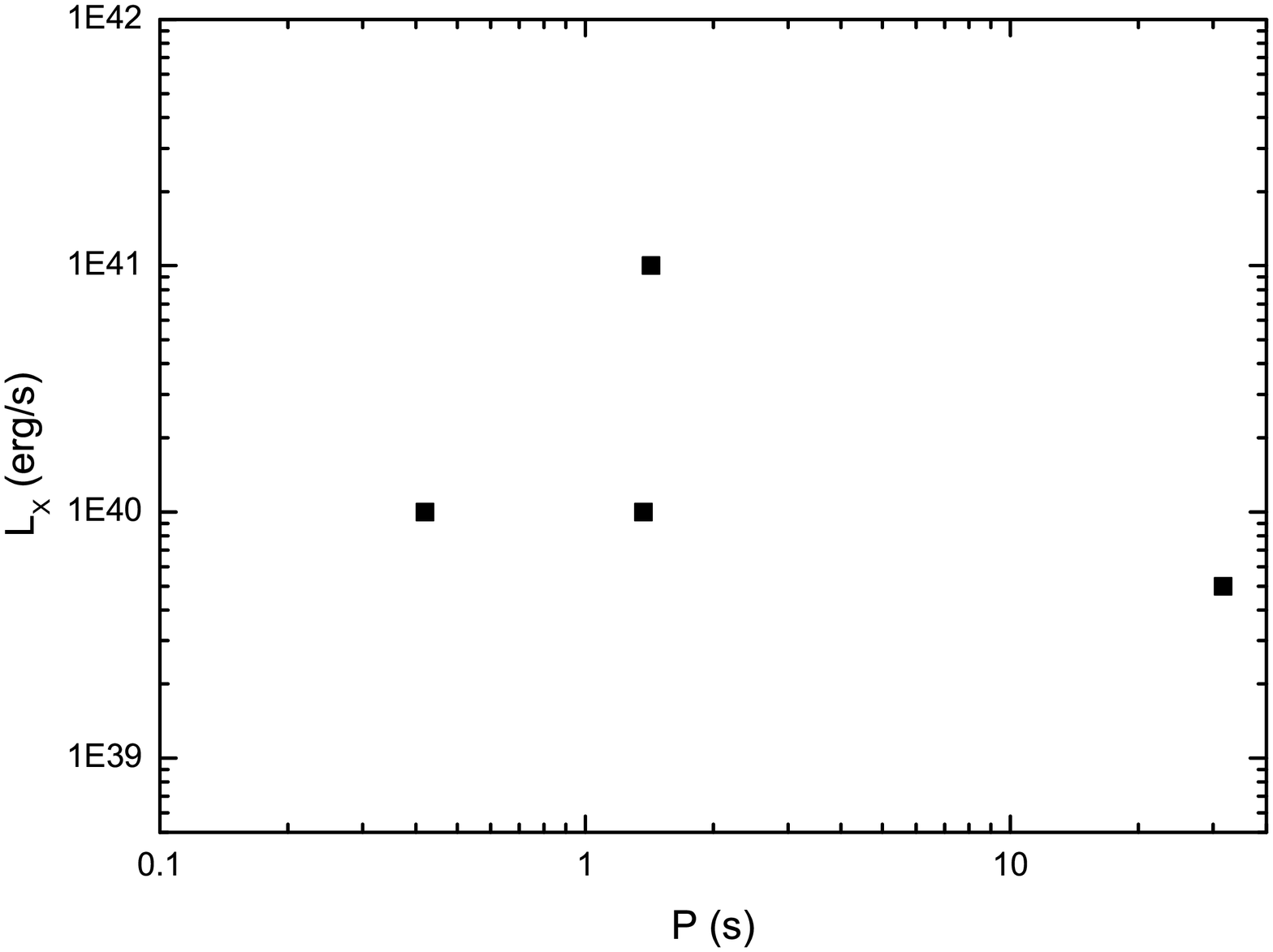}
\end{minipage}
\begin{minipage}{0.45\textwidth}
 \includegraphics[width=0.95\textwidth]{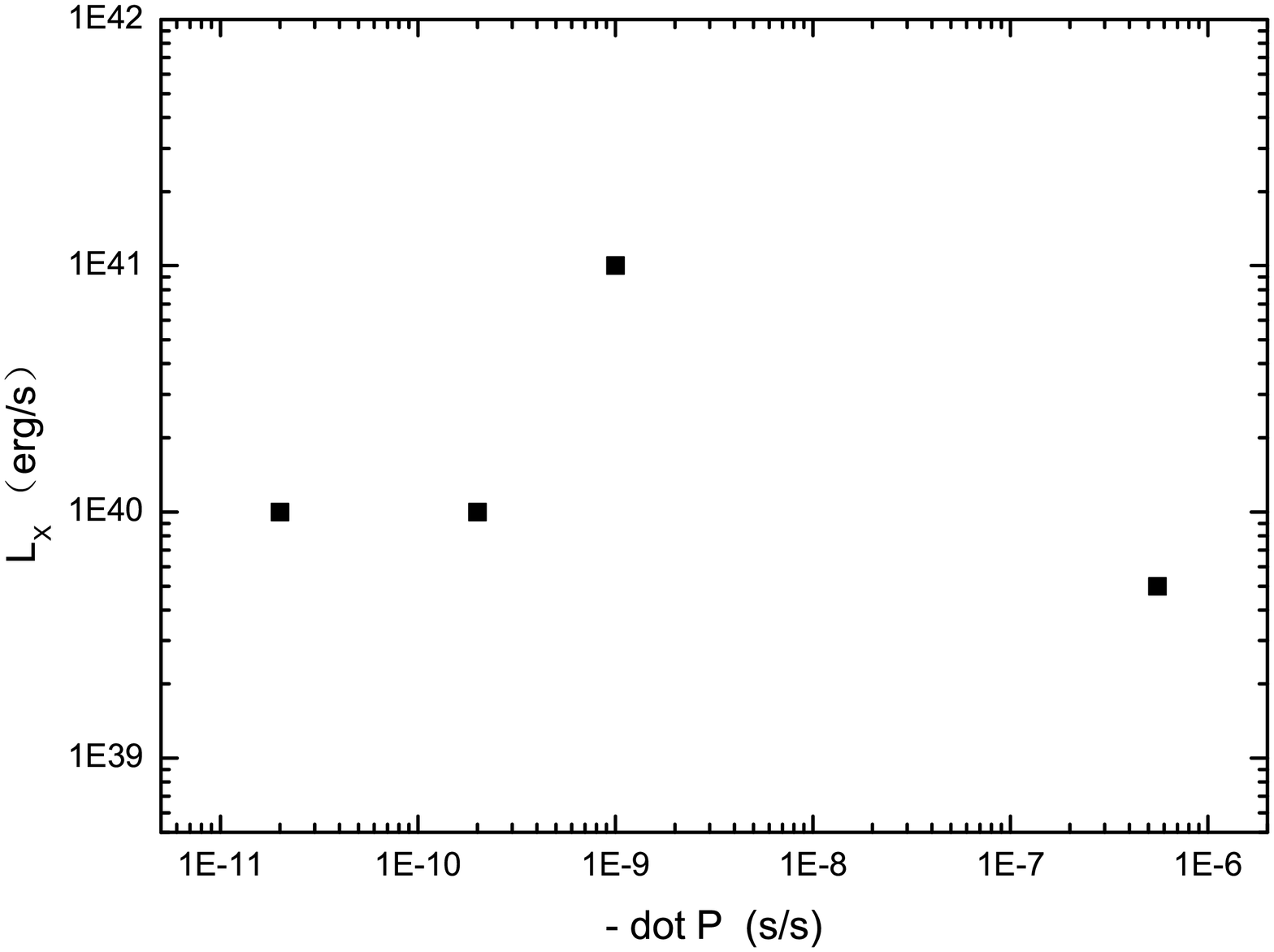}
\end{minipage}
\begin{minipage}{0.45\textwidth}
 \includegraphics[width=0.95\textwidth]{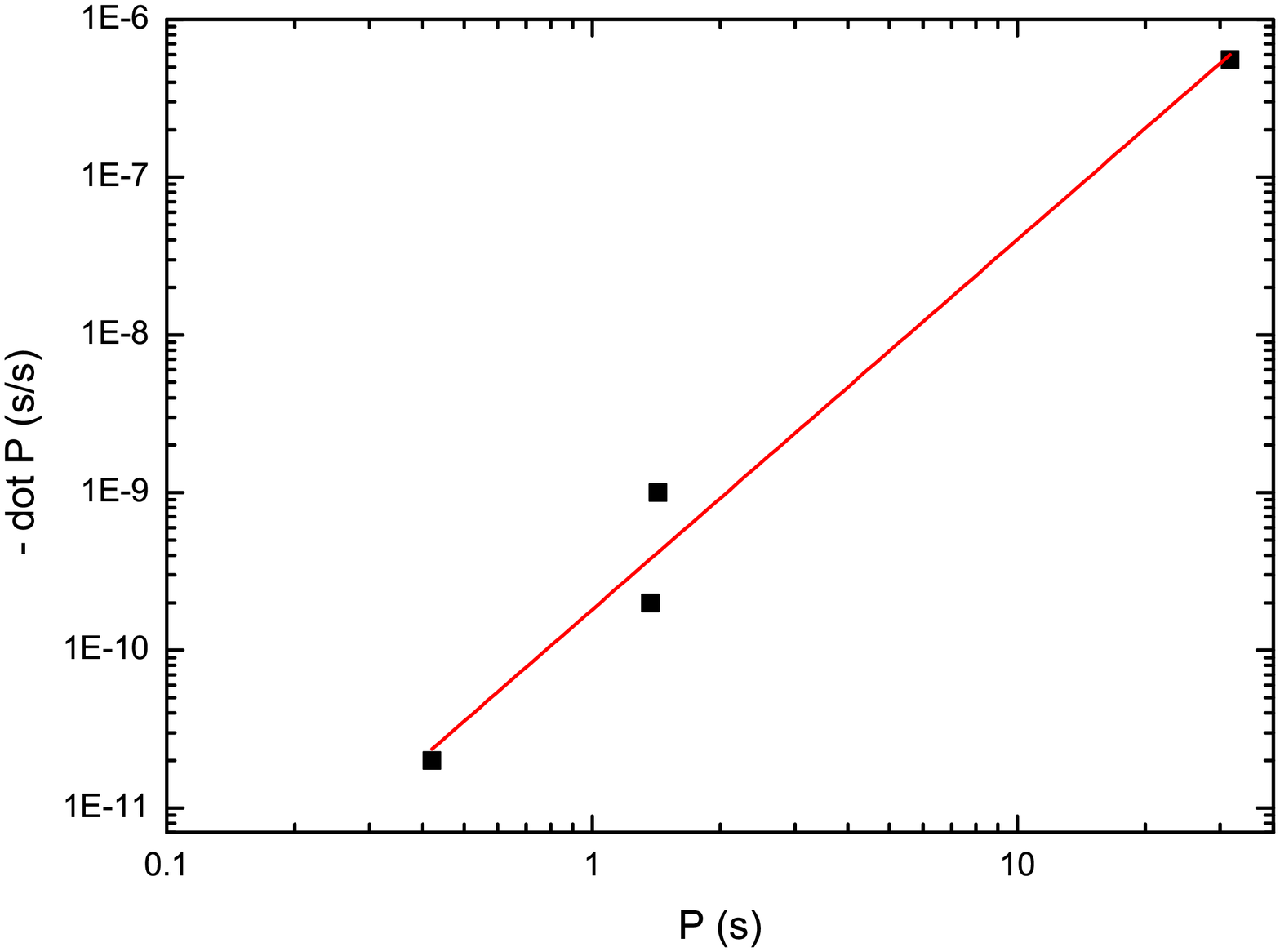}
\end{minipage}
%
\caption{The diagrams of the observed three parameters in ULX pulsars: $P$, $\dot{P}$ and $L_{\rm x}$. In the top panel, we also plot the diagram of $-\dot{P}$ versus $(P L_{\rm x}^{3/7})^2$. }
\label{fig_ppdot}
\end{figure}

In equation (\ref{eqn_spinup}), considering that the Alfv$\rm \acute{e}$n radius is proportional to $\dot{M}^{-2/7}$, therefore
\begin{equation}
\dot{\Omega} \propto \dot{M}^{6/7} \propto L_{\rm x}^{6/7}.
\end{equation}
In terms of pulsation period and period derivative (Ghosh \& Lamb 1979; Shapiro \& Teukolsky 1983)
\begin{equation}
-\dot{P} \propto (P L_{\rm x}^{3/7})^2.
\end{equation}
This correlation between spin-up rate and X-ray luminosity can be revealed in two aspects: (1) For a large number of ULX pulsars,
such a correlation may be seen. (2) For a single source, variations in its X-ray luminosity will result in variation of its spin-up rate;  a higher luminosity will result in a higher spin-up rate (F$\rm \ddot{u}$rst  et al. 2016). Possible correlations of this kind have been found in normal accreting neutron stars (Sugizaki et al. 2017). In Figure \ref{fig_ppdot}, the diagrams of three parameters $P$, $\dot{P}$ and $L_{\rm x}$ of the four ULX pulsars are plotted.  There exists a strong correlation between $P$ and $\dot{P}$ for the known ULX pulsars. The red solid line in Figure \ref{fig_ppdot} shows the fitted best relation between $P$ and $\dot{P}$: $\dot{P} \propto P^{2.3\pm 0.2}$. For the four ULX pulsars, we cannot find the correlations of $L_{\rm x}$ versus $P$ or  $L_{\rm x}$ versus $\dot P$. The diagram of $-\dot P$ and the parameter $(P L_{\rm x}^{3/7})^2$ is also plotted, which shows a general trend of a positive correlation.

\subsection{General constraint on the mass accretion rate, beaming factor and dipole magnetic field}

In the above calculations, a specific beaming factor is assumed. Some general constraints on the mass accretion rate, beaming factor and dipole magnetic field can be obtained for ULX pulsars. The following calculations are motivated by Chen (2017)'s calculation for M82 X-2. Dall'Osso et al. (2015) also obtained similar limits for M82 X-2 assuming the accretion torque of Ghosh \& Lamb (1979). Comparison between our results and those of Dall'Osso et al. (2015) will be presented following each item.

During the outburst state, ULX pulsars are spinning up due to accretion. The spin-up torque may be dominated by the matter torque, equation (\ref{eqn_spinup}). For accretion to occur, the Alfv$\rm \acute{e}$n radius should be smaller than the corotation radius
\begin{equation}\label{eqn_accretion}
R_{\rm A} \le R_{\rm co} \equiv \left(  \frac{G M}{4\pi^2} \right)^{1/3} P^{2/3}.
\end{equation}
By combining equation (\ref{eqn_spinup}) and (\ref{eqn_accretion}), a lower limit on the mass accretion rate can be obtained
\begin{equation}
\dot{M} \ge \dot{M}_{\rm min} = 3.6\times 10^{18} \left( \frac{\dot{P}}{-10^{-10}} \right) \left( \frac{P}{1\,\rm s} \right)^{-7/3} \,\rm g \ s^{-1},
\end{equation}
where typical values of period and period-derivative are inserted. Using the timing observations of the four ULX pulsars, their minimum mass accretion rates are: $3.4\times 10^{18} \,\rm g \,s^{-1}$ (for M82 X-2), $9.4\times 10^{18} \,\rm g \,s^{-1}$ (for NGC7793 P13), $1.3\times 10^{20} \,\rm g \,s^{-1}$ (for NGC5907 ULX pulsar),
and $6.3\times 10^{18} \,\rm g\, s^{-1}$ (for NGC300 ULX1). This confirms their super-Eddington nature.

Using the lower limit for the mass accretion rate, a lower limit on the beaming factor can be obtained:
$L_{\rm x} = b L_{\rm x,iso} =
\eta \dot{M} c^2 \ge \eta \dot{M}_{\rm min} c^2$. The lower limit on the beaming factor is
\begin{eqnarray}
b \ge b_{\rm min} &=& \frac{\eta \dot{M}_{\rm min} c^2}{L_{\rm x,iso}}\\ \nonumber
&=&0.03 \frac{\eta}{0.1} L_{\rm x,iso,40}^{-1} \left( \frac{\dot{P}}{-10^{-10}} \right) \left( \frac{P}{1\,\rm s} \right)^{-7/3}.
\end{eqnarray}
For the four ULX pulsars, the lower limits on their beaming factors are: $0.03 (\eta/0.1) L_{\rm  x,iso,40}^{-1}$ (for M82 X-2), $ 0.08 (\eta/0.1) L_{\rm x,iso,40}^{-1}$ (for NGC7793 P13), $0.12 (\eta/0.1) L_{\rm x,iso,41}^{-1}$ (for NGC5907 ULX pulsar), and $0.12(\eta/0.1)(L_{\rm x,iso}/4.7\times 10^{39} \,\rm erg \, s^{-1})^{-1}$ (for NGC300 ULX1).
In the previous section, a beaming factor of $b=0.2$ is assumed. It is consistent with the lower limits obtained here. In some calculations, a strong beaming factor is assumed. The lower limits here are not in strong support for this scenario.

Dall'Osso et al. (2015)'s lower limits on mass accretion rate of M82 X-2 is twice as large as ours because they used a specific accretion torque\footnote{The function $n(\omega_s)\omega_s^{1/3}$ has a maximum value of 0.5 in Dall'Osso et al. (2015), where $n(\omega_s)$ is the dimensionless torque, $\omega_s$ is the fastness parameter. See section 2.6 below for more details.}, while we used the matter torque, equation (\ref{eqn_spinup}). Dall'Osso et al. (2015)'s lower limit on the beaming factor of M82 X-2 is a factor of four larger than ours, because they used an energy conversion efficiency\footnote{The symbol $\eta$ has different meanings in Dall'Osso et al. (2015).} about $0.2$ instead of our choice of $\eta=0.1$ and their $\dot{M}_{\rm min}$ is twice as large as ours.

Furthermore, considering that the beaming factor should be smaller than one $b\le 1$, an upper limit on the mass accretion rate can be obtained. From $\eta \dot{M} c^2 = b L_{\rm x,iso} \le L_{\rm x,iso}$, then
\begin{equation}
\dot{M} \le \dot{M}_{\rm max}= \frac{L_{\rm x,iso}}{\eta c^2} =10^{20} \left(\frac{\eta}{0.1} \right)^{-1} L_{\rm x,iso,40}  \rm \ g \ s^{-1}.
\end{equation}
For M82 X-2, NGC7793 P13,  and NGC300 ULX1, the upper limit on their mass accretion is about $10^{20} \rm \ g \ s^{-1}$. For NGC5907 ULX pulsar, its isotropic X-ray luminosity can be as high as $10^{41} \rm \ erg \ s^{-1}$. The upper limit on its mass accretion is about $10^{21} \rm \ g \ s^{-1}$. Dall'Osso et al. (2015)'s upper limits on the mass accretion rate of M82 X-2 is about $0.6\times 10^{20} \ \rm g \ s^{-1}$. It is slightly lower than ours. The difference is mainly caused by a different choice of energy conversion efficiency, as stated above.

From equation (\ref{eqn_accretion}), an upper limit on the dipole magnetic field as a function of mass accretion rate can be obtained
\begin{equation} \label{Bmax}
B_{\rm p} \le B_{\rm p,max} =2.2\times 10^{12} \dot{M}_{18}^{1/2} P^{7/6} \rm \ G,
\end{equation}
where $\dot{M}_{18}$ is the mass accretion rate in units of $10^{18} \rm \ g \ s^{-1}$.
For a specific ULX pulsar, its mass accretion rate is limited by $\dot{M}\le \dot{M}_{\rm max}$.  A conservative limit on the maximum magnetic field can be obtained. For M82 X-2, it is about $3\times 10^{13} \ \rm G$. Dall'Osso et al. (2015)'s upper limit for M82 X-2 is about $2\times 10^{13} \ \rm G$. Their value is slightly smaller than ours, because their $\dot{M}_{\rm max}$ is slightly smaller than ours, which is again due to a different choice of energy conversion efficiency.
The difference is that equation (\ref{eqn_accretion}) is independent on the specified accretion torque. And the resulting upper limit on magnetic field (equation (\ref{Bmax})) is a function of the mass accretion rate.

Since pulsation can be detected from the ULX pulsars, their Alfv$\rm \acute{e}$n radius should be larger than the
neutron star radius $R_{\rm A} \ge R$. This means that the neutron star dipole magnetic field can not be arbitrarily small.
This places a lower limit on the neutron star dipole magnetic field (as a function of mass accretion rate)
\begin{equation}\label{Bmin}
B_{\rm p} > B_{\rm p,min}= 2.8\times 10^8 \dot{M}_{\rm 18}^{1/2} \ \rm G.
\end{equation}
During the calculations, a typical neutron star radius of $10\ \rm km$ is assumed\footnote{The dependence on
the neutron star mass and radius is: $B_{\rm p,min} \propto (M/1.4 \ M_{\odot})^{1/4} (R/10\rm \ km)^{7/4}$}.
This idea is similar to that of accreting normal neutron stars (Bildstein et al. 1997; Zhang \& Kojima 2006). Accreting neutron stars in high mass
X-ray binaries are often observed as X-ray pulsars. While, the pulsation of accreting neutron stars in low mass X-ray binaries are
often not detected. This may because neutron stars in low mass X-ray binaries have a lower dipole magnetic field. This also means
that if some accreting magnetars have a low dipole magnetic field, then they may not been observed as pulsating sources. Finding pulsations is only one way to confirm the neutron star nature of ULX sources.
Dall'Osso et al. (2015)'s lower limits on the mass accretion rate is about $5\times 10^9 \ \rm G$, corresponding to an accretion rate $\dot{M} \sim 38 \times 10^{18} \ \rm g \ s^{-1}$ (inset in their Figure 2). According to equation (\ref{Bmin}), the corresponding lower limit at that accretion rate is about $1.7 \times 10^9 \ \rm G$. A factor of three difference is caused by the coefficient of $0.5$ in the definition of the magnetospheric radius. The lower limits on magnetic field here (equation (\ref{Bmin})) is a function of the corresponding mass accretion rate. The accretion torque information is not required in obtaining equation (\ref{Bmin}). While Dall'Osso et al. (2015) chose a specific accretion torque and combined it with equation(\ref{Bmin}). Therefore, a specific numerical value for the magnetic field lower limit can be obtained there. 

For every ULX pulsar, they have both lower limit and upper limit on the mass accretion rate. Therefore, the upper/lower limits of their dipole magnetic field can be plotted as a function of their mass accretion rate. From equation (\ref{eqn_spinup}), the neutron star dipole magnetic field can be found as a function of the true mass accretion rate
\begin{equation}\label{eqn_B}
B_{\rm p} = 2\times 10^{14} \dot{M}_{18}^{-3} \left( \frac{\dot{P}}{-10^{-10}} \right)^{7/2} \left( \frac{P}{1\rm \, s} \right)^{-7}
\rm \ G.
\end{equation}
Considering the super-Eddington nature of ULX pulsars, their mass accretion rate can be very high $\dot{M} \gg 10^{18} \rm \ g \ s^{-1}$. This will result in $B_{\rm p} \ll 2\times 10^{14} \rm \ G$. This is in general consistent with the accreting low magnetic field magnetar scenario.

\begin{figure}
\begin{minipage}{0.45\textwidth}
 \includegraphics[width=0.95\textwidth]{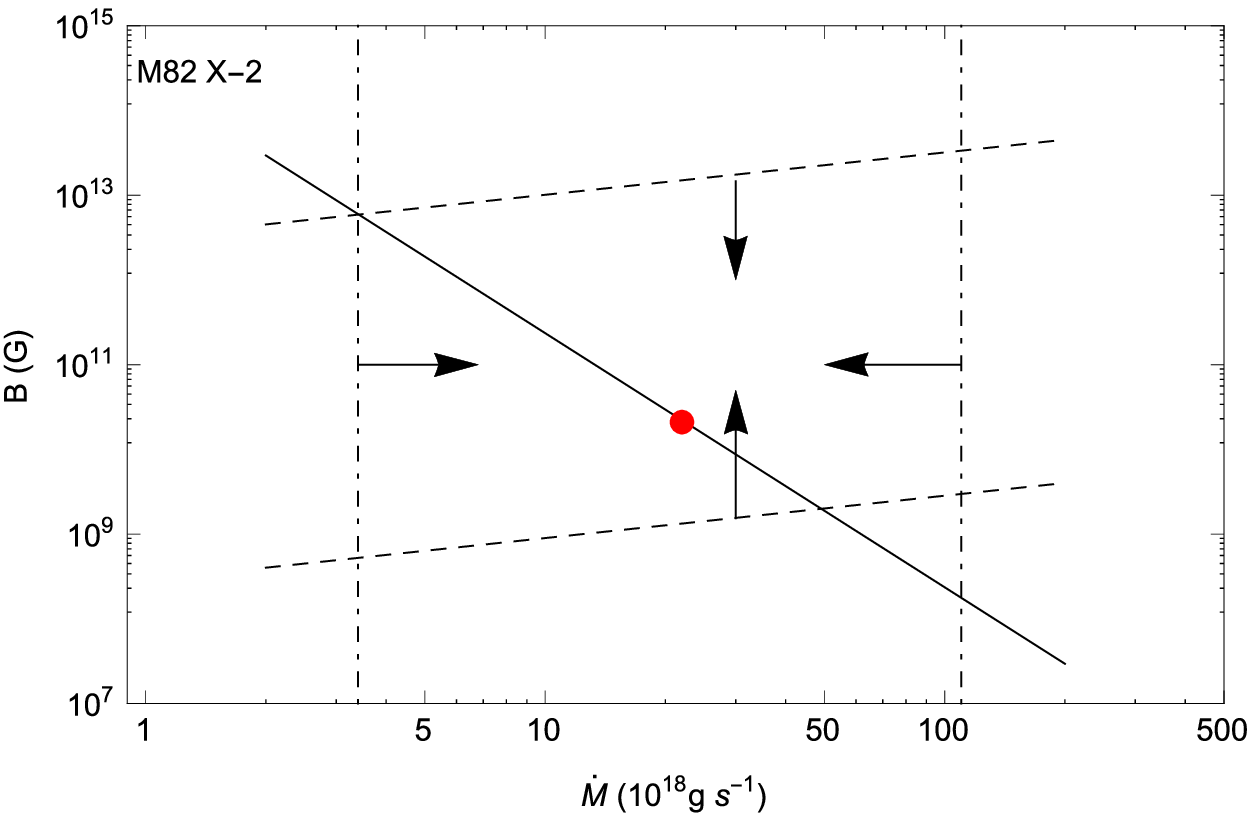}
\end{minipage}
\begin{minipage}{0.45\textwidth}
 \includegraphics[width=0.95\textwidth]{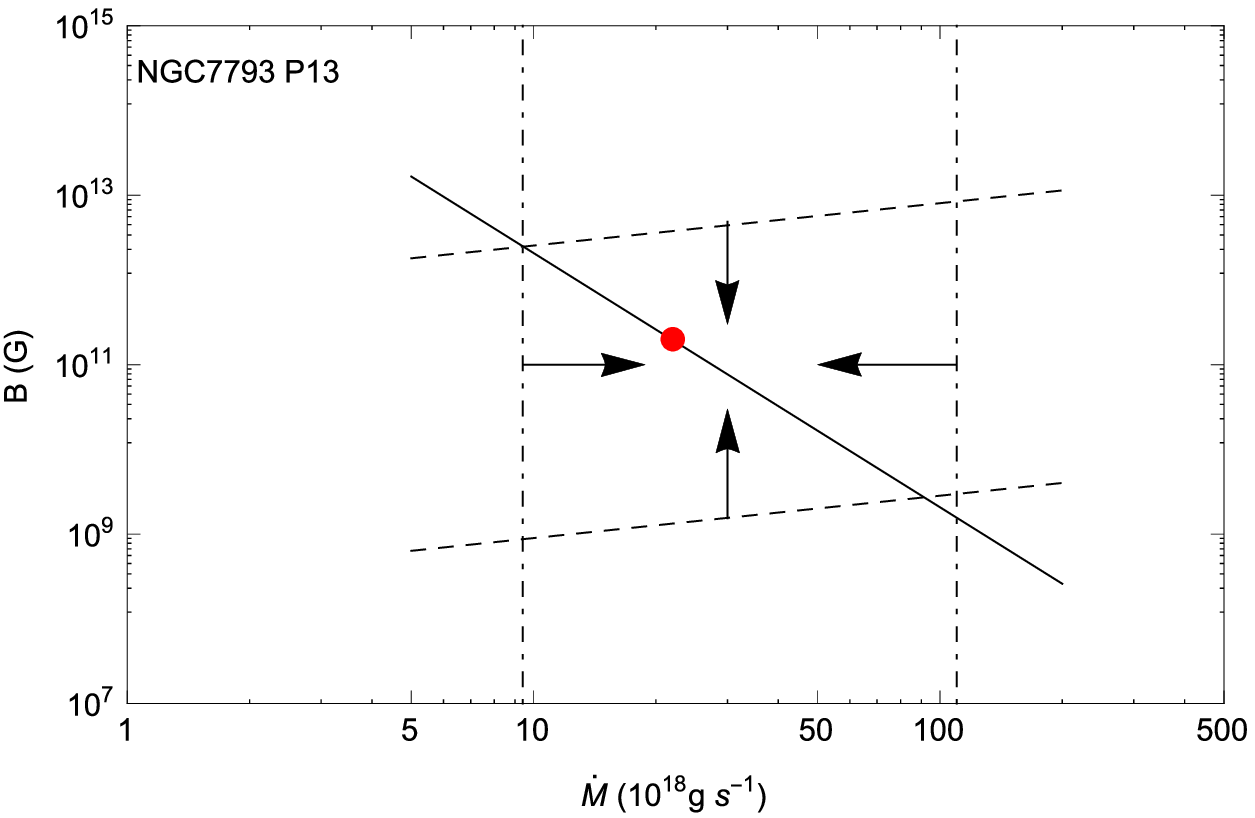}
\end{minipage}
\begin{minipage}{0.45\textwidth}
 \includegraphics[width=0.95\textwidth]{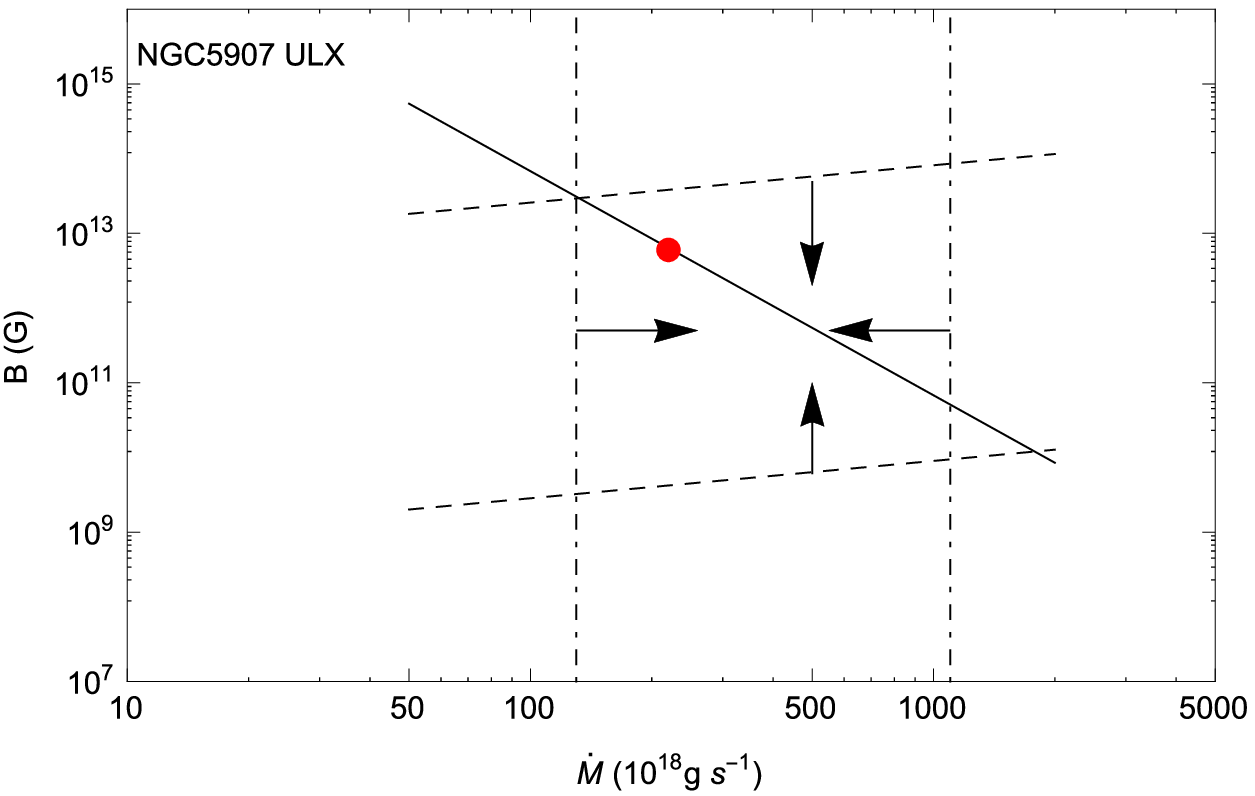}
\end{minipage}
\begin{minipage}{0.45\textwidth}
 \includegraphics[width=0.95\textwidth]{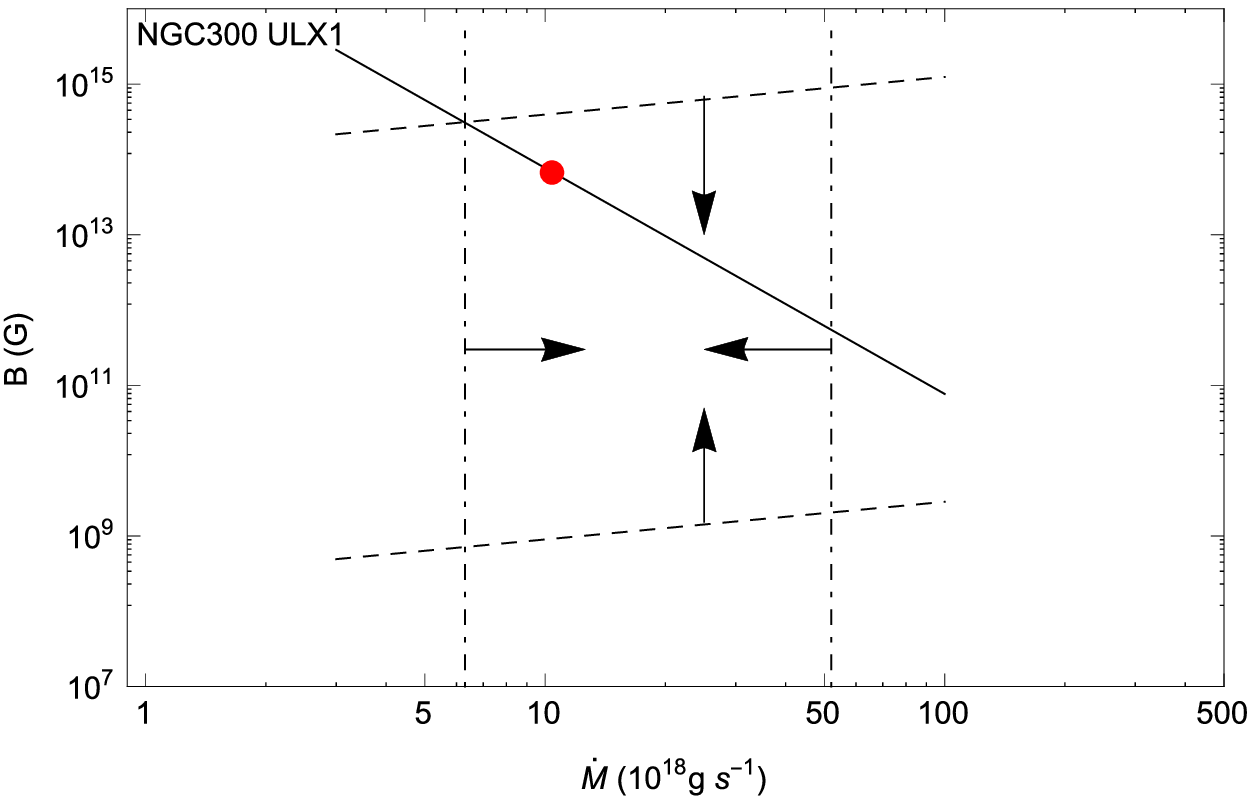}
\end{minipage}
\caption[]{Dipole magnetic field as a function of mass accretion rate of the four ULX pulsars. The source name is shown in the top left corner in each panel. The vertical dot-dashed lines
are the lower/upper limits of the ULX pulsar's mass accretion rate. The dashed lines are the upper/lower limits of the dipole magnetic field.
The solid line is the dipole magnetic field in equation (\ref{eqn_B}).  The red point corresponds to the
calculations in section \ref{section_B_calculation}, assuming typical values of beaming factor and accretion efficiency.}
\label{fig_BMdot}
\end{figure}

Figure \ref{fig_BMdot} shows the dipole magnetic field vs. mass accretion rate of ULX pulsars. In Figure \ref{fig_BMdot}, the region between the two vertical dot-dashed lines and the two dashed lines is the allowed range of mass accretion rate and dipole magnetic field. In this region, the solid line is the locus of possible solutions according to equation (\ref{eqn_B}).
The red point corresponds to a specific choice for the beaming factor, as discussed in section 2.2.

\subsection{Duty cycle of the ULX state and accretion equilibrium}

Observationally, the ULX pulsars switch between the high luminosity state and the low luminosity state. This may be due to the transition of the neutron star between the accretion phase and the propeller phase (Tong 2015; Dall'Osso et al. 2015, 2016; Tsygankov et al. 2016). X-ray luminosity observations may tell us some information of the duty cycle of the ULX state (high luminosity state). A finite duty cycle of the ULX state will also affect the timing behavior of the neutron star in two aspects.
\begin{enumerate}
\item The neutron star will spin up during the ULX state, and spin down  during the low luminosity state. Over the long term, the neutron star may approach an equilibrium period. This will be more physical compared with defining a rough long term averaged mass accretion rate (equation (\ref{eqn_Peq})).

\item The neutron star's measured spin-up torque will be larger during the ULX state than its long term averaged value. This has already been found in observations (Israel et al. 2017b).
\end{enumerate}

A unified torque for the accretion phase and the propeller phase may be employed (Menu et al. 1999; Tong et al. 2016)
\begin{equation}
N \propto \dot{M} R_{\rm A}^2 \Omega_{\rm K} (R_{\rm A}) \left( 1- \frac{\Omega}{\Omega_{\rm K}(R_{\rm A})} \right),
\end{equation}
where $\Omega_{\rm K} (R_{\rm A})$ is the Keplerian angular velocity at the Alfv$\rm \acute{e}$n radius. Denoting the the accretion rate during the ULX state (high state) as $\dot{M}_{\rm h}$, its duty cycle is defined as $D_{\rm p}$. Denoting the mass accretion rate during the low state as $\dot{M}_{ l}$, its duty cycle will be $1-D_{\rm p}$. The rotational evolution of the neutron star is governed by $N=I \dot{\Omega}$. Accretion equilibrium means that: $\int N dt = I \Delta \Omega=0$, therefore
\begin{eqnarray}
 \nonumber
&&\dot{M} R_{\rm A}^2 \Omega_{\rm K} (R_{\rm A}) \vert_{\dot{M} = \dot{M}_{\rm h}} D_{\rm p}\\
&&- \dot{M} R_{\rm A}^2 \Omega_{\rm K} (R_{\rm A}) \frac{\Omega}{\Omega_{\rm K}(R_{\rm A})} \vert_{\dot{M} = \dot{M}_{ l}} (1-D_{\rm p}) =0.
\end{eqnarray}
Solving for the duty cycle
\begin{equation}
\frac{1-D_{\rm p}}{D_{\rm p}}
= \frac{\dot{M} R_{\rm A}^2 \Omega_{\rm K} (R_{\rm A}) \vert_{\dot{M} = \dot{M}_{\rm h}}}{\dot{M} R_{\rm A}^2 \Omega \vert_{\dot{M} = \dot{M}_{ l}}}.
\end{equation}
Using that the Alfv$\rm \acute{e}$n radius is proportional to $\dot{M}^{-2/7}$ (equation (\ref{eqn_Alfven radius})), then
\begin{eqnarray}
\frac{1-D_{\rm p}}{D_{\rm p}} & =& \left( \frac{\dot{M}_{\rm h}}{\dot{M}_{ l}} \right)^{3/7}
\frac{\Omega_{\rm K}(R_{\rm A}) \vert_{\dot{M} = \dot{M}_{\rm h}}}{\Omega}\\
&=& \left( \frac{\dot{M}_{\rm h}}{\dot{M}_{ l}} \right)^{3/7} \frac{1}{\omega_{\rm s}},
\end{eqnarray}
where $\omega_{\rm s} =\Omega/\Omega_{\rm K}(R_{\rm A})\vert_{\dot{M} = \dot{M}_{\rm h}}$ is the fastness
parameter during the ULX state. In the accretion state, the fastness parameter is smaller than one: $\omega_{\rm s}<1$.
Then
\begin{equation}
\frac{1-D_{\rm p}}{D_{\rm p}} > \left( \frac{\dot{M}_{\rm h}}{\dot{M}_{ l}} \right)^{3/7}.
\end{equation}
Further simplification of the above equation shows that
\begin{equation}
D_{\rm p} < \left( \frac{\dot{M}_{ l}}{\dot{M}_{\rm h}} \right)^{3/7}.
\end{equation}
For typical values of $\dot{M}_{ l} \sim 10^{18} \ \rm g \ s^{-1}$ and $\dot{M}_{\rm h} \sim 10^{20} \ \rm g \ s^{-1}$,
the upper limit on the duty cycle of the ULX state (especially for the peak luminosity) is $D_{\rm p} < 14\%$. This limit on the duty cycle of the ULX state is roughly consistent
with the X-ray observations (Tsyganov et al. 2016; F$\rm \ddot{u}$rst  et al. 2016; Israel et al. 2017a,b).

Considering that the propeller torque is very effective, then for most of the time the neutron star will be in accretion equilibrium during the low state with accretion rate $\dot{M}_{l}$. In the outburst state with accretion rate $\dot{M}_{h}$, the neutron star will spin up and try to run from the previous equilibrium state to the new equilibrium state. The net effect of spin-up during the outburst state will be balanced by the spin-down during the low state. For the above typical parameters, the low state accretion rate is 100 times lower than that of the high state. However, with a duty cycle in the low state several times larger than that of the high state, the effect of $\dot{M}_{l}$ can balance that of $\dot{M}_{h}$. This is because the lever arm (i.e., Alfv$\rm \acute{e}$n radius) in the low state is larger than that in the high state.

\subsection{Which equilibrium and path to equilibrium}

\begin{figure}
\begin{minipage}{0.45\textwidth}
 \includegraphics[width=0.95\textwidth]{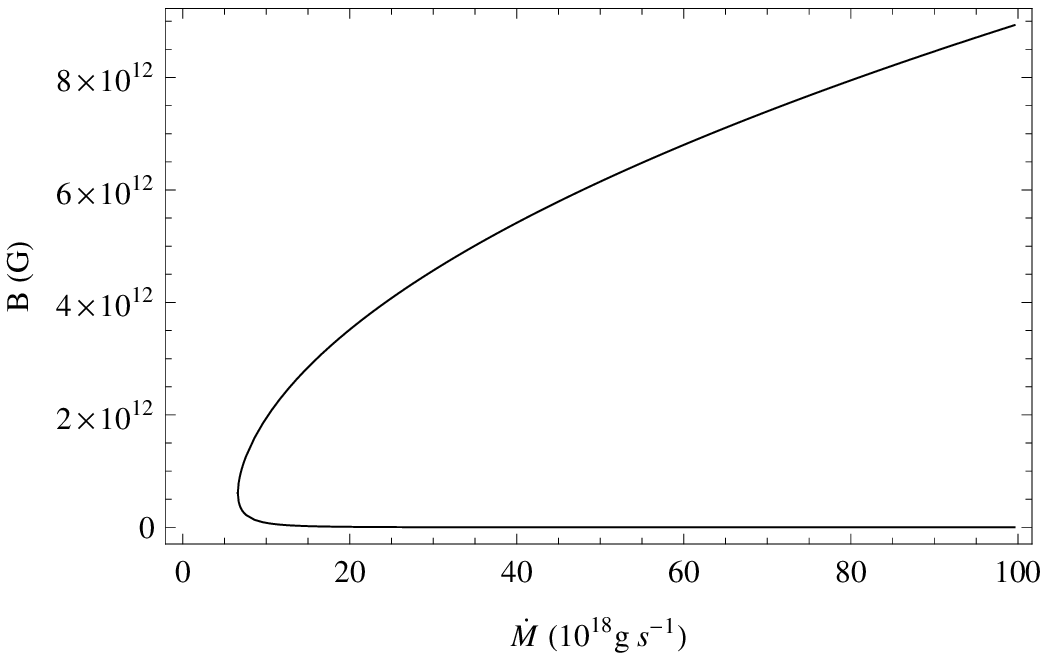}
\end{minipage}
\begin{minipage}{0.45\textwidth}
 \includegraphics[width=0.95\textwidth]{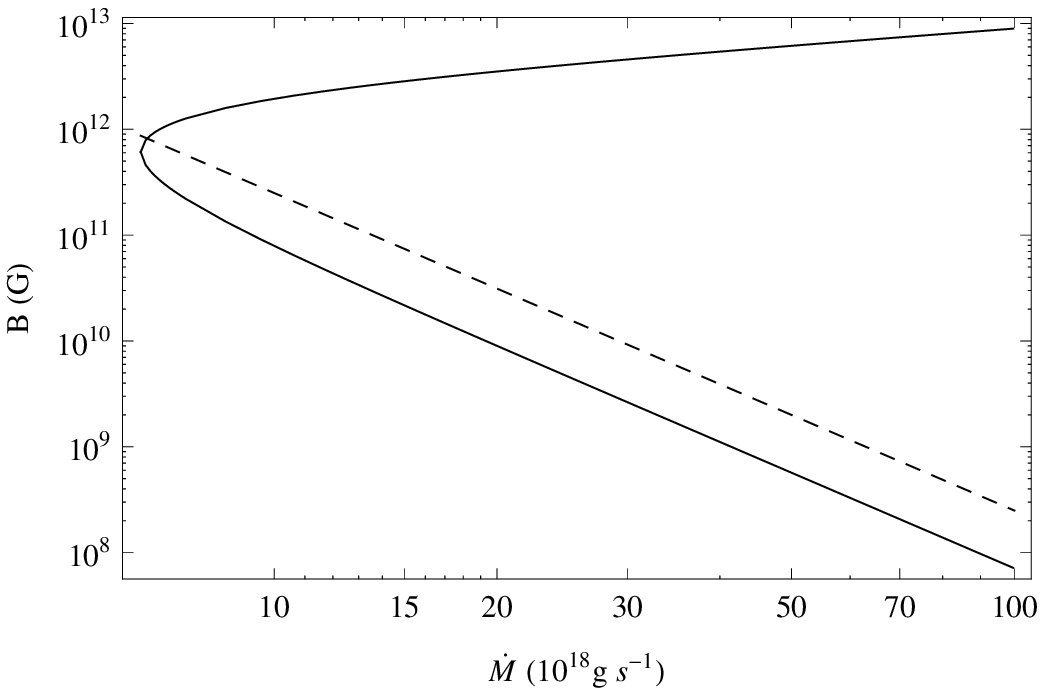}
\end{minipage}
\caption[]{The dipole magnetic field of M82 X-2 as a function of mass accretion rate, assuming a general accretion torque.
The upper panel is in linear scale, the lower panel is in logarithmic scale. The dashed line in lower panel is equation (\ref{eqn_B}), which is obtained from equation (\ref{eqn_spinup}). }
\label{fig_BMdot_general}
\end{figure}

In Dall'Osso et al. (2015), a general accretion torque is considered (Ghosh \& Lamb 1979). According to Dall'Osso et al. (2015), a dipole magnetic field about $10^{13} \ \rm G$ is preferred for M82 X-2, and the neutron star should be near accretion equilibrium.
In Ghosh (1995),  a simplified form of the accretion torque of Ghosh \& Lamb (1979) is provided
\begin{equation}\label{eqn_general_torque}
n(\omega_s) = 1.4 \left( \frac{1-\omega_s/\omega_c}{1-\omega_s} \right),
\end{equation}
where $n(\omega_s)$ is the dimensionless torque function, $\omega_s =\Omega/\Omega_{\rm K} (R_{\rm A})$ is the fastness parameter, and $\omega_c \approx 0.35$ is the critical fastness.  The total torque is the matter torque (equation (\ref{eqn_spinup})) times the dimensionless torque: $ \dot{M} \sqrt{G M R_{\rm A}} n(\omega_s)$. We have made calculations using both the original torque form of Ghosh \& Lamb (1979), and the simplified torque form of Ghosh (1995). The results are almost the same. Therefore, the simplified torque of Ghosh (1995) will be employed.
If equation (\ref{eqn_spinup}) is replaced by the general accretion torque (equation (\ref{eqn_general_torque})), the magnetic field of M82 X-2 can also be found as a function of the mass accretion rate, see Figure \ref{fig_BMdot_general}.

Figure \ref{fig_BMdot_general} is the same as Figure 2 in Dall'Osso et al. (2015), except for some differences in the definitions of magnetospheric radius, accretion efficiency etc.
As discussed in Dall'Osso et al. (2015), the solution consists of an upper branch and a lower branch. If the accretion rate is assumed to be equal to that during the outburst, the upper branch is near accretion equilibrium. The lower branch is dominated by the matter torque. For slow rotators, $\omega_s \ll 1$, the dimensionless torque becomes $n(\omega_s) \approx 1.4$. The total torque is 1.4 times of that in equation (\ref{eqn_spinup}). This causes the differences between the dashed line and the lower branch in Figure \ref{fig_BMdot_general}.

If in the outburst state, the neutron star is in accretion equilibrium, then the magnetic field in the upper branch should be chosen. This is the conclusion of Dall'Osso et al. (2015).
For accreting neutron stars, it is possible that they are in accretion equilibrium in the long run. And the long term averaged mass accretion rate may be much smaller than that during the outburst. During the outburst, they should be in the slow rotator regime. Therefore, the magnetic field in the lower branch may be preferred. This is the difference between our scenario and that of Dall'Osso et al. (2015).

One consequence of the above scenario is that: if the neutron star is in accretion equilibrium in the long run, then during the outburst it will run from the previous equilibrium state to the new equilibrium state. During the outburst state, the mass accretion rate is higher, the corresponding equilibrium period becomes shorter. Then the neutron star will spin up. For a $\dot{M}_{h}$ one hundred times larger than $\dot{M}_l$, the equilibrium period in the outburst state will be seven times smaller (see equation (\ref{eqn_Peq})). The NGC300 ULX1 pulsar shows a rapid period evolution: the neutron star may approaching its equilibrium period asymptotically (Carpano et al. 2018). The period evolution may be fitted with a constant term plus an exponential decay term (Carpano et al. 2018). This asymptotical behavior may be demonstrated analytically under some assumptions.  Assuming a constant accretion rate during the outburst, and a simplified accretion torque $n(\omega_s) =1-\omega_s$, then the angular velocity will evolve with time as
\begin{equation}
\dot{\Omega} = \frac{\dot{M} \sqrt{G M R_{\rm A}}}{I} \left( 1-  \frac{\Omega}{\Omega_{\rm K}(R_{\rm A})} \right).
\end{equation}
In terms of pulsation period, it will evolve with time as
\begin{equation}\label{eqn_P_evolution}
\dot{P} = - \frac{\dot{M} \sqrt{G M R_{\rm A}}}{2\pi I} P^2 \left(  1- \frac{P_{\rm eq}}{P} \right).
\end{equation}
Early in the outburst, at time $t_i$, the neutron star pulsation period is denoted as $P_i$. Then the pulsation period at time $t$ can be found by integrating equation (\ref{eqn_P_evolution})
\begin{equation}
P(t) = \frac{P_{\rm eq}}{ 1- \left(  1- \frac{P_{\rm eq}}{P_i} \right) e^{- \frac{t-t_i}{\tau}} },
\end{equation}
where $\tau = (\dot{M} \sqrt{G M R_{\rm A}}/(2\pi I) \times  P_{\rm eq})^{-1} = I/(\dot{M} R_{\rm A}^2)$ is a typical time scale. The period evolution with time depends on the equilibrium period $P_{\rm eq}$ and the time scale $\tau$. These two parameters are determined by the accretion rate and magnetic field of the neutron star. Therefore, by fitting the period evolution with time, the accretion rate and magnetic field may be determined simultaneously.  For an initial pulsation period much larger than the equilibrium period $P_i \gg P_{\rm eq}$ and $t-t_i \gg \tau$, the period evolution can be further simplified as: $P = P_{\rm eq} (1+e^{-(t-t_i)/\tau})$. The analytical solutions here are consistent with
the rapid period evolution of NGC300 ULX1 pulsar (Carpano et al. 2018). If the accretion torque contains a critical fastness, e.g. $n(\omega_s) = 1-\omega_s/\omega_c$, then the critical fastness can be absorbed in the definition of the equilibrium period.
If the accretion is of more general case (or more complicated case), the period evolution with time may be more complicated.

\section{Accreting magnetar scenario for slow pulsation X-ray pulsars}

The X-ray source AX J1910.7$+$0917 has pulsation period of $36.2\ {\rm ks} \sim 10 \ \rm h$ (Sidoli et al. 2017).
Its X-ray luminosity ranges from $1.7\times 10^{34} \ \rm erg \ s^{-1}$-$10^{36} \ \rm erg \ s^{-1}$.
It may be the slowest pulsation X-ray pulsar at present. The nature of AX J1910.7$+$0917 may be a wind accreting neutron star
with normal magnetic field about $10^{12} \,\rm G$ (Sidoli et al. 2017 and references therein). According to the
equilibrium period of disk accretion (equation (\ref{eqn_Peq})), if AX J1910.7$+$0917 has a dipole magnetic field
of $4\times 10^{15} \ \rm G$ (corresponds to $\mu_{30} =2\times 10^3$) and accretion rate of $\dot{M}_{17} \sim 10^{-3}$,
then its pulsation period can also be understood in the accreting magnetar scenario. It has a very long pulsation period
because its magnetar strength dipole magnetic field and a low accretion rate. The variation of its X-ray luminosity
may be due to variation of the mass accretion rate.  

Observations of other slow pulsation X-ray pulsars may give some indirect support for the accreting magnetar nature of  AX J1910.7$+$0917.
\begin{itemize}
\item The long period neutron star  4U 0114$+$65 (with rotational period about $9500\ \rm s$) may be an accreting magnetar candidate (Sanjurjo-Ferrrin et al. 2017). Observationally, possible signatures of a transient disk were also found in this system (Hu et al. 2017).

\item The central neutron star inside the supernova remnant RCW 103 may have a rotational period about $6.7$ hours (De Luca et al. 2006). Magnetar-like activities were found in this source, confirming its magnetar nature (Rea et al. 2016; D'Ai et al. 2016).
The physical reason for its long rotational period may be accretion from a fallback disk (Tong et al. 2016).  Therefore, the long period neutron star inside RCW 103 is a magnetar accreting from a fallback disk.
\end{itemize}
The spin-down torque due to a disk is very efficient. If AX J1910.7$+$0917 has an initial rotational period similar to that of isolated magnetars (about $10\rm \ s$) at the beginning of accretion, with dipole magnetic field of $4\times 10^{15} \rm \ G$, and mass accretion rate $10^{14} \rm \ g \ s^{-1}$, then it can evolve to a rotational period about $\sim 30\ \rm ks$ in a very short time\footnote{It is because the neutron star has a braking index of one and its rotational period increases exponentially in the propeller phase. For example, adopting the propeller torque in Tong et al. (2016), the spin-down time scale is about 40 years. }.
Similar to the case of 4U 0114$+$65, the possible disk in AX J1910.7$+$0917 may be a transient accretion disk (i.e., the disk has a duty cycle smaller than one.). Since the disk torque is very efficient, the transient disk may still dominate the rotational behavior of the central neutron star.

There are several merits  of understanding the slowest X-ray pulsar AX J1910.7$+$0917  in the accreting magnetar scenario.
\begin{enumerate}
\item  If ULX pulsars are accreting magnetars with mass accretion rate about $10^{20} \rm g \ s^{-1}$, then there can also be accreting magnetars with a lower mass accretion rate. The slowest X-ray pulsar with low luminosity and long period may correspond to this case.

\item The prediction for AX J1910.7$+$0917 is very clear if it is an accreting magnetar. With a magnetic field about $10^{15} \rm G$, and a low mass accretion rate $\dot{M} \sim 10^{14} \rm g \ s^{-1}$, inside the Galaxy, magnetar-like activities may also been observed in this source.

\item Different kinds of special accreting neutron stars may be linked together in the accreting magnetar scenario. (a) ULX pulsars may be accreting magnetars with high mass accretion rates. (b) The slow pulsation X-ray pulsars may be accreting magnetars with low mass accretion rates, including AX J1910.7$+$0917, 4U 0114$+$65 (Sanjurjo-Ferrrin et al. 2017), 4U 2206$+$54 (Reig et al. 2012) etc. Some supergiant fast X-ray transients (SFXTs, e.g. sources which can have very low X-ray luminosities, Walter et al. 2015) may be accreting magnetars transit between the accreting phase and the propeller phase (Bozzo et al. 2008). (c) Accreting magnetars with intermediate mass accretion rates ($10^{17}-10^{18} \rm g \ s^{-1}$), and relative long spin period ($\sim 1000 \rm \ s$) may correspond to slow pulsation X-ray pulsars in the Small Magellanic Cloud (Klus et al. 2014; Ho et al. 2014). In Figure 3, the spin period and orbital period of the known ULX pulsars, SFXT pulsars and slow pulsation pulsars in the Galaxy and the Small Magellanic Cloud are collected.
\end{enumerate}

\begin{figure}
\centering
\includegraphics[width=0.55\textwidth]{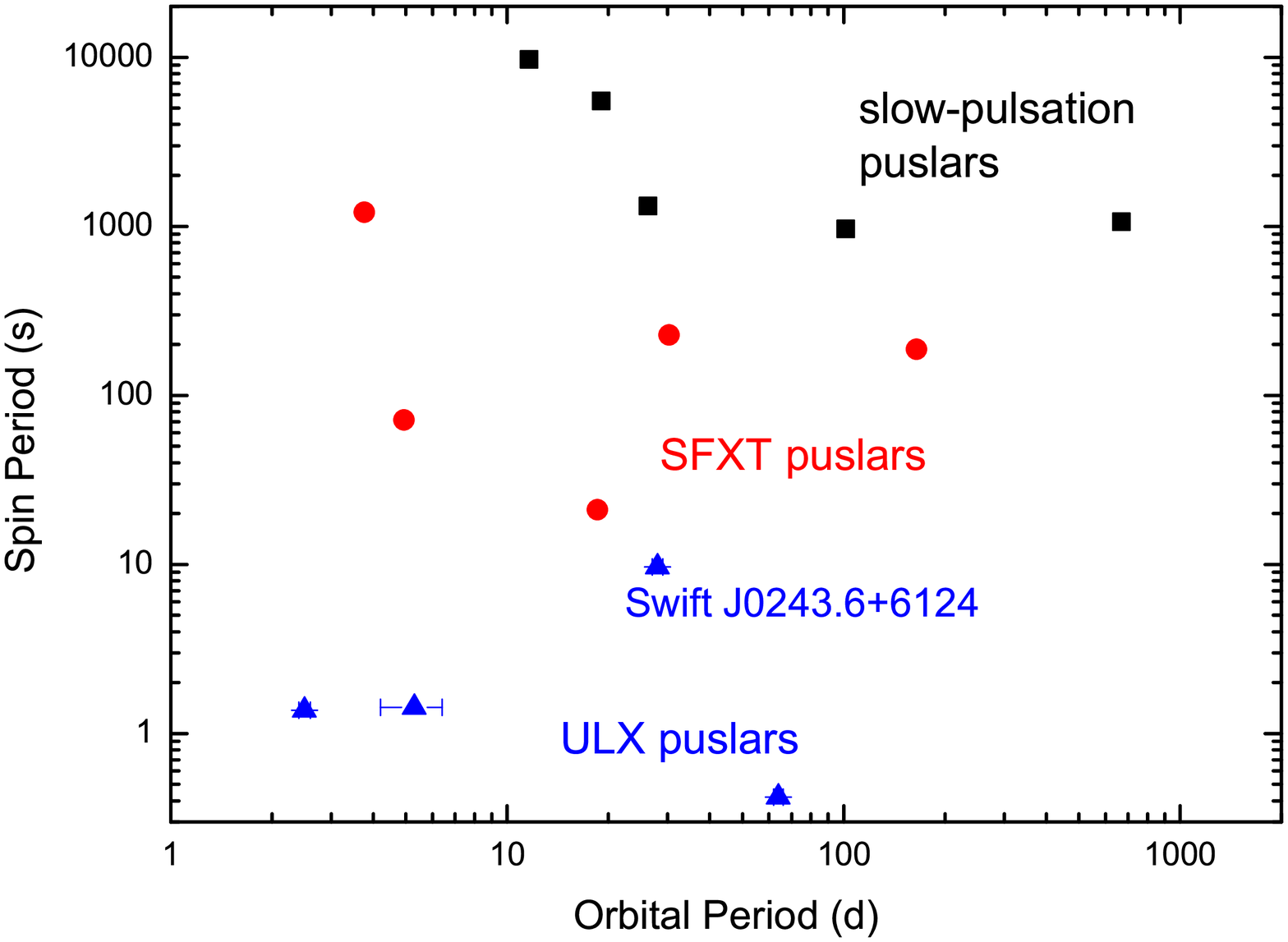}
\caption{The diagram of the spin period versus orbital period in ULX pulsars, SFXT pulsars and slow pulsation X-ray pulsars in the Galaxy and the Small Magellanic Cloud. The data points of ULX pulsars (including the Galactic ULX pulsar candidate Swift J0243.6+6124) are taken from Bachetti et al. (2014), Israel et al. (2017a,b), Kennea et al. (2017) and Doroshenko et al. (2018); the data points of SFXTs are taken from Sidoli (2017); the data points of the slow pulsation X-ray pulsars are taken from Klus et al. (2014), Wang (2009, 2011), Gonzalez-Galan et al. (2018).}
\label{fig_spin_orbit}
\end{figure}

ULX pulsars may have a very high mass accretion rate. The accreted matter may result in a decay of the magnetar's magnetic field (Pan et al. 2016). It is similar to the accretion induced magnetic field decay in low mass X-ray binaries  (Shibazaki et al. 1989; Zhang \& Kojima 2006). This may explain why ULX pulsars tend to be accreting low magnetic field magnetars (besides possible magnetic field decay without accretion, Israel et al. 2017b). Other accreting magnetar candidates do not have such high mass accretion rate, the magnetic field decay there may be not be very significant. This may be the first difference between ULX pulsars and other accreting magnetar candidates. Wind accretion may have difficulties in explain the very high X-ray luminosity of ULX pulsars. Therefore, the second difference is that ULX pulsars accretes via Roche lobe overflow (Bachetti et al. 2014), while slow pulsation X-ray pulsars are wind accreting systems. There are also links between ULX pulsars and slow pulsation X-ray pulsars. The possible transient disks in wind accreting neutron stars (Koh et al. 1997; Jenke et al. 2012; Hu et al. 2017) help to link these two kinds of accreting magnetar candidates. In Figure \ref{fig_spin_orbit}, it is suggestive that longer orbital period may result in a longer rotational period of the neutron star for these accreting magnetar candidates.

Previously the two outstanding slow pulsation X-ray pulsars are 4U 0114+65 (or 2S 0114+65) and 4U 2206+54. 4U 2206+54 has a rotational period about $5600\ \rm s$ and is spinning down at a rate about $5\times 10^{-7} \rm s\ s^{-1}$ (Reig et al. 2012; Wang 2013). 4U 0114+65 has a rotational period about $9500 \ \rm s$ and is spinning up at a rate $-1\times 10^{-6} \rm  s\ s^{-1}$ (Wang 2011; Hu et al. 2017).  It was proposed that there may be evolutions between these two systems, and ``it is possible that there exists a candidate neutron star binary with a spin period much longer than 10 000 s which may be the product of the long-term spin down of the 4U 2206+54-like neutron star and would undergo the spin-up transition to form the
2S 0114+65-like source'' (see the caption of Figure 10 in Wang 2013).  We think that  AX J1910.7$+$0917 with a pulsation period of $3.6\times 10^4 \rm \ s$ may be the missing link between 4U 2206+54 and 4U 0114+65. The period derivative of  AX J1910.7$+$0917 has not been determined at present (Sidoli et al. 2017). Adopting the typical value of 4U 2206+54 and 4U 0114+65, an absolute value of period derivative can be estimated to be $\sim 10^{-6} \rm s \ s^{-1}$ (either spin-up or spin-down). The rotational period of  AX J1910.7$+$0917 is determined to be $36200\pm 110 \rm \ s$ at the year of $2011$ (Sidoli et al. 2017).  In about 30 years thereafter, the rotational period of  AX J1910.7$+$0917 will change by an amount of $\sim 1000\rm \ s$. Timing observations of this source in the not very far future can test this possibility.

\section{Discussions}

Accreting magnetars may have different mass accretion rate and different manifestations.
However, confirming the central neutron star's magnetar nature is not easy, since a lack of direct measurement of the neutron star's magnetic field (Revnivtsev \& Mereghetti 2015). There are many possible signatures of an accreting magnetar. A summary is listed in below.
\begin{enumerate}
\item Magnetar-like outbursts. Similar things have already happened in confirming the magnetar nature of the RCW 103 central neutron star (Rea et al. 2016; D'Ai et al. 2016).  However, it is not known at present whether and to what degree accretion will affect the magnetic activities
of the central magnetar. At least in the case of low mass accretion rate, magnetic activities may still be present. Therefore, the slowest X-ray pulsar AX J1910.7$+$0917, and other long period X-ray pulsars may show some magnetar-like activities in the future.

\item A hard X-ray tail. Isolated magnetars tend to have a high X-ray tail above $100\,\rm keV$ compared with accreting normal neutron stars (Mereghetti et al. 2015). The possible signature of a hard X-ray tail in 4U 2206$+$54 and 4U 0114+65 may also hint a magnetar nature of the central neutron star (Reig et al. 2012; Wang 2011, 2013). This hard X-ray tail may be of magnetic origin (like isolated magnetars) or due to interaction of the soft X-ray photons with the accretion flow in strong magnetic field.

\item ULX pulsars. As in the case of magnetar giant flares (Paczynski 1992), a highly super-Eddington luminosity may be due to the presence of a magnetar strength magnetic field. The magnetic field concerned is the total magnetic field at the neutron star surface. While, the one used in torque studies is mainly the large scale dipole field.

\item Cyclotron line observations (if the final magnetic field is in the magnetar range). However, whether the line is due to electron or proton origin may be uncertain (Brightman et al. 2018; Walton et al. 2018b). The conclusion will be very different when assuming different origins.

\item Switch between the accretion phase and propeller phase (if the final magnetic field is in the magnetar range). By figuring a specific luminosity corresponding to the transition, the magnetic field can be determined (Tsygankov et al. 2016; Dall'Osso et al. 2016).

\item ULX sources with pulsar-like spectra.  Some sources may have small pulsed fraction due to interaction between the neutron star's dipole magnetic field and the accretion flow. Pulsar-like spectra may indicate a neutron star at the center (Pintore et al. 2017; Walton et al. 2018a).

\item Slow puslation X-ray pulsars. A slow pulsation combined with a low X-ray luminosity may indicate an accreting magnetar with low mass accretion rate. The slowest pulsation X-ray pulsar AX J1910.7$+$0917, besides other slow pulsation X-ray pulsars, may be such candidates.

\end{enumerate}

\section*{Acknowledgments}

We thank the referee for constructing criticisms.
H. Tong is supported by the NSFC (11773008). W. Wang is supported the National Program on Key Research
and Development Project (Grants No. 2016YFA0400803) and the NSFC (11622326).


\label{lastpage}

\end{document}